\documentclass[10pt,english,aip,reprint,twocolumn,floatfix]{revtex4-2}
\usepackage[T1]{fontenc}
\usepackage[latin9]{inputenc}
\usepackage{color}
\usepackage{amsmath}
\usepackage{amssymb}
\usepackage{graphicx}
\usepackage{revsymb} 

\usepackage[colorlinks, linkcolor=blue, anchorcolor=red, citecolor=blue]{hyperref}
\usepackage{graphicx}
\usepackage{bm}

\usepackage{mathptmx}
\usepackage{etoolbox}  

\usepackage{babel}

\begin{document}

\global\long\def\d{\mathrm{d}}%
\global\long\def\bx{\boldsymbol{x}}%
\global\long\def\bp{\boldsymbol{p}}%
\global\long\def\bv{\boldsymbol{v}}%
\global\long\def\bq{\ensuremath{\boldsymbol{q}}}%
\global\long\def\bk{\boldsymbol{k}}%
\global\long\def\be{\boldsymbol{e}}%

\title{\bfseries\sffamily Validation of Classical Transport Cross Section for Ion-Ion   
    Interactions Under Repulsive Yukawa Potential}

\author{Tian-Xing Hu}
\affiliation{Key Laboratory for Laser Plasmas and School of Physics and Astronomy,
and Collaborative Innovation Center of IFSA, Shanghai Jiao Tong University,
Shanghai, 200240, China}
\author{Dong Wu}\email{dwu.phys@sjtu.edu.cn}
\affiliation{Key Laboratory for Laser Plasmas and School of Physics and Astronomy,
and Collaborative Innovation Center of IFSA, Shanghai Jiao Tong University,
Shanghai, 200240, China}
\author{C. L. Lin}
\affiliation{National Key Laboratory of Computational Physics, Institute of Applied 
Physics and Computational Mathematics, Beijing 100088, People's Republic of China.}
\affiliation{Key Laboratory for Laser Plasmas and School of Physics and Astronomy,
and Collaborative Innovation Center of IFSA, Shanghai Jiao Tong University,
Shanghai, 200240, China}
\author{Z. M. Sheng}
\affiliation{Institute for Fusion Theory and Simulation, Department of Physics,
Zhejiang University, Hangzhou 310027, China.}
\author{B. He}\email{he\_bin@iapcm.ac.cn}
\affiliation{National Key Laboratory of Computational Physics, Institute of Applied 
Physics and Computational Mathematics, Beijing 100088, People's Republic of China.}
\author{J. Zhang}
\affiliation{Key Laboratory for Laser Plasmas and School of Physics and Astronomy,
and Collaborative Innovation Center of IFSA, Shanghai Jiao Tong University,
Shanghai, 200240, China}
\affiliation{Institute of Physics, Chinese Academy of Sciences, Beijing 100190, China.}

\begin{abstract}
    Value of cross section is a fundamental parameter to depict the transport of charged particles in matters.
    Due to masses of orders of magnitude higher than electrons and convenience of realistic calculation, 
    the cross section of elastic nuclei-nuclei collision is usually treated via classical mechanics. 
    The famous Bohr criterion was firstly proposed to judge whether the treatment via classical mechanics is reliable or not. 
    Later, Lindhard generalized the results of Coulomb to screening potentials.
    Considering the increasing importance of detailed ion-ion interactions under modern simulation codes in inertial confinement fusion (ICF) researches,
    the validation of classical transport cross section for ion-ion   
    interactions in a big range of parameter space is certainly required. 
    In this work, the transport cross sections via classical mechanics under repulsive Yukawa potential are compared 
    with those via quantum mechanics. 
    Differences of differential cross sections are found with respect to scattering angles and velocities. 
    Our results generally indicate that the classical picture fails at the cases of both low and high velocities, 
    which represent a significant extension of the famous Bohr criterion and its generalized variations.
    Furthermore, the precise validation zones of classical picture is also analysed in this work. 
    This work is of significant importance for benchmarking the modern ion-kinetic simulation codes in ICF researches, 
    concerning the stopping power of $\alpha$ particles in DT fuels, ion-ion friction and viscous effects in the formation of kinetic shocks.    
\end{abstract}

\maketitle

\section{Introduction}
Ion-ion collision is of fundamental importance in inertial confinement fusion (ICF) researches \citep{Montgomery2018,Lin_2023,Fu2020,Wang2014}.
For example, at the latter stage of ignition and burning ware propagation, more than a half of the energy of fusion $\alpha$ particles is deposit directly into DT ions
through ion-ion collision \citep{brown_charged_2005}.
Especially, at the end of range, the dominant stopping power of $\alpha$ particles comes from ion-ion interactions, 
in which both deflections and decelerations need to be taken seriously. 
In terms of modern ion-kinetic simulation code, accurate modelling of ion-ion collisions is vital, 
for example, the Monte-Carlo collision method widely used in PIC or hybrid-PIC code \cite{Wu2020,Zhang2020,Chen2020} require an explicit and ``easy-to-handle'' ion-ion transport cross section. 

In most of the literatures, the ion-ion collision is treated via classical approaches.
Due to the increasing importance of detailed ion-ion interactions under modern ion-kinetic simulation codes in ICF researches,
the validation of classical transport cross section for ion-ion   
interactions in a big range of parameter space is certainly required.

Transport cross-section (TCS) is the basis in order to model ion-ion collisions \cite{sigmund2006particle}. 
For matters at room temperature or plasmas, the TCS is usually calculated based on integration of 
two classical particles \cite{bonderup1981penetration}, where quantum mechanical effects are often neglected. 

In the pioneering work of N. Bohr \cite{bohr1948penetration}, a criterion on the applicability
of classical orbital picture in the calculation of Coulomb scattering 
was proposed, and it is called the Bohr criterion nowadays.
The Bohr criterion was then generalized by Lindhard to the case of the screened field such as the standard atomic potential
 \cite{lindhard1965influence} in 1965, and to the case of Yukawa potential later \cite{sigmund2006particle}. 
Both Bohr and Lindhard's works were based on small angles approximations and suggested that the classical orbital picture fail at high velocities.
However the contribution of all the scattering angles must be included to get the TCS. 

TCS is usually defined as 
\begin{equation}
\sigma_{\mathrm{tr}}(v)=\int(1-\cos\theta)\sigma(\theta,v)\d\Omega \label {ctrdc},
\end{equation} 
where $\sigma(\theta,v)$ is the differential cross-section (DCS) with $\theta$ the scattering angle at the frame of centre of mass when the relative velocity between the projectile 
and the target particle is $v$. 
In most of the literatures and articles \cite{le2023hybrid,Wu2017,Wu2020}, 
when it comes to ion-ion collision, the classical method is
often just taken for granted without any justification. 
To our knowledge,
only E. Bonderup \cite{bonderup1981penetration} offered a brief analysis of this issue based 
on the Linhard standard potential for small-angle scattering. Bouderup qualitatively concluded that the classical 
orbital picture fails at large collision parameter or small scattering angle, at which the contribution 
to stoppig power is negligible, but did not give a quantitative comparison. 
Moreover, he also pointed out that this is true for any other potentials. 

The repulsive Yukawa potential is widely used to describe the ion-ion interaction in plasma physics researches.
The calculation of classical TCS under Yukawa potential is much simpler than that of quantum TCS,
the reliability of which is very important to the research of the transportion of charged particles. 
Despite being a fundamental problem, the quantitatively comparison between the quantum and classical 
result is currently lacking.The main purpose of this paper is to discuss the applicability or
the scope of application of Bohr's criterion to the TCS due to ion ion collison  under Yukawa potential in detail.
It is found that the classical collision picture is failed at both very low and high velocity cases. 
In the work we will take the example of the collisons of different charge state ions with DT ion to study the validity of classical TCS. 
Some numerical results are presented for a few screening lengths of the Yukawa potential since no analytical expressions can be found for the potential. 

The paper is organized as following. In Sec.\ \ref{secMethod},
we briefly introduce the classical and quantum mechanical methods
in the calculation of TCS involved in this paper. In Sec.\ \ref{sec:Applicability-of-the} and IV,
we use these numerical methods to test the validity of the generalized
Bohr's criterion for DCS and TCS, respectively. And finally the conclusions are presented in the last section. The atomic units
are used throughout the work unless otherwise explicitly indicated.

\section{Method}\label{secMethod}
Firstly, we briefly review the classical and quantum mechanical methods
for calculating the DCS and TCS. In the work the potential for binary collision is always the Yukawa potential 
  $U(r)=\frac{Z_{1}Z_{2}}{r}\mathrm{e}^{-r/\lambda_{0}}$.  Here $\lambda_{0}$ is the characteristic screening length. 
 And $Z_1$ and $Z_2$ ($Z_1 Z_2 > 0$) are the charges for the projectile and the target ion, respectively. 

\subsection{Classical Scattering}
The classical scattering angle under the potential $U(r)$
is given by
\begin{equation}
    \theta(b)=\pi-2b\int_{r_{0}}^{\infty}\frac{\d r/r^{2}}{\sqrt{1-b^{2}/r^{2}-U(r)/\mathcal{E}}},\label{classical scattering angle}
\end{equation}
where $b$ is the collisional parameter, $r_{0}$ is the apsis of the
scattering, and $\mathcal{E}=\mu v^2/2$, with $\mu$ being the effective mass $m_{1}m_{2}/\left(m_{1}+m_{2}\right)$, is the incident energy. 
 Here $m_1$ and $m_2$ are the masses for the projectile and the target ion, respectively.
With the change
of variable \cite{ORAIFEARTAIGH1983255}

\begin{equation}
    u=\sqrt{\frac{r_{0}}{r}-1},
\end{equation}
Eq.\ (\ref{classical scattering angle}) become
\begin{equation}
    \theta(b)=\pi-4b\int_{0}^{1}\frac{\d u}{\sqrt{g(u)}},
\end{equation}
where
\begin{equation}
    g(u)=b^{2}\left(2-u^{2}\right)+\frac{r_{0}^{2}}{u^{2}\mathcal{E}}\left[U(r_{0})-U\left(\frac{r_{0}}{1-u^{2}}\right)\right].
\end{equation}
Thus, the singularity at the apsis $r_{0}$ is avoided. The differential
cross-section (DCS) is then
\begin{equation}
    \sigma(\theta)=\frac{b}{\sin\theta}\left|\frac{\d b}{\d\theta}\right|.\label{cdc}
\end{equation}

\subsection{Quantum Scattering}

We will introduce three major quantum mechanical methods involved
in this paper. Firstly, the Born approximation is the perturbation method
for quantum scattering processes.  We consider only the first-order approximation in this paper, 
and the result serves as a benchmark of our numerical results in high-velocity limit. 
Secondly, the partial wave method (PWM)
is the standard approach which calculates the DCS/TCS by expanding the scattering wave into a series of 
spherical harmonics and then solve the phase shifts order by order. Since the scattering amplitude would 
eventually tends to zero provided the quantum number $\ell$ is large enough, we can obtain a result
with any satisfactorily accuracy by means of the PWM.
Lastly, the WKB approximation is an semi-classical method of the wave equation i.e., the 
approximation of geometric optics, which is easier to calculate at both middle and high velocity regime than the exact PWM.

\subsubsection{Born Approximation}

For the Yukawa potential
the scattering
cross-section calculated via Born approximation $\sigma_{\mathrm{Born}}^{\mathrm{Y}}(\theta)$, which is only valid
in high velocity limit, is related to the Rutherford expression
\begin{equation}
    \sigma^{\mathrm{C}}(\theta)=\frac{b_{0}^{2}}{4\sin^{4}\frac{\theta}{2}}
\end{equation}
by
\begin{equation}
    \sigma_{\mathrm{Born}}^{\mathrm{Y}}(\theta)=\sigma^{\mathrm{C}}(\theta)\left(1+\frac{1}{q^{2}\lambda_{0}^{2}}\right)^{-2},
    \label{born_app}
\end{equation}
where
\begin{equation}
    q=2k\sin\frac{\theta}{2}\ \text{and}\ b_{0}=\frac{Z_{1}Z_{2}}{\mu v^{2}},
\end{equation}
are the momentum transfer for each collision and the collision radius
(which represents the $90^{\circ}$ deflection impact parameter for
classical coulomb collision) respectively. And $k=\mu v$ 
stands for the momentum of the collision system. From these the total cross section is 
\begin{equation}
    \sigma_\mathrm{tot}=\frac{16\pi\mu^2\lambda_0^2Z_1^2Z_2^2}{1+4\mu^2 v^2\lambda_0^2},
\end{equation}
When $v$ is high enough, it becomes $4\pi\lambda_0^2Z_1^2Z_2^2/v^2$.

\subsubsection{Partial Wave Method }

The exact solution of the phase shift obeys the differential equation
\cite{calogero1967variable}

\begin{equation}
    \frac{\d\delta_{\ell}}{\d r}=-\frac{2\mu U}{k}\left[\hat{j}_{\ell}(kr)\cos\delta_{\ell}-\hat{n}_{\ell}(kr)\sin\delta_{\ell}\right]^{2},\label{deltaleq}
\end{equation}
and the scattering amplitude is

\begin{equation}
    f(\theta)=\sum_{\ell=0}f_\ell(\theta)=\frac{1}{k}\sum_{\ell=0}^{\infty}(2\ell+1)e^{i\delta_{\ell}}\sin\delta_{\ell}P_{\ell}(\cos\theta).\label{f}
\end{equation}
and
\begin{equation}
    \sigma(\theta)=\left|f(\theta)\right|^{2}.\label{qdc}
\end{equation}
One can calculate the transport cross-section directly by the phase
shift \cite{mora2020coulomb}:

\begin{equation}
    \sigma_{\mathrm{tr}}(v)=\frac{4\pi}{k^{2}}\sum_{\ell=0}^{\infty}(\ell+1)\sin^{2}\left(\delta_{\ell}-\delta_{\ell+1}\right),\label{qtrdc}
\end{equation}
instead of by Eq.\ (\ref{ctrdc}). Certainly it is impossible to
consider the infinite number of partial waves, however, in practice,
only a few of them contribute to very low velocity
collisions.

\subsubsection{Approximation Method}

As the relative velocity $v$ increases, more and more partial waves should
be taken into consideration, and it is difficult to solve equation
(\ref{deltaleq}) for high $\ell$ waves. In such cases, the WKB approximation
is therefore a very good approximation \cite{rodberg1968introduction}:

\begin{equation}
    \begin{aligned}\delta_{\ell}^{\mathrm{WKB}}= & \int_{r_{1}}^{\infty}\d r\sqrt{k^{2}-\frac{(\ell+1/2)^{2}}{r^{2}}-2\mu U(r)} \\
                              & -\int_{r_{0}}^{\infty}\d r\sqrt{k^{2}-\frac{(\ell+1/2)^{2}}{r^{2}}},
    \end{aligned}
\end{equation}
where $r_{0}$ and $r_{1}$ are the apsis of free and scattering particles respectively.

On the other hand, for very large $\ell$, the Legendre function $P_\ell(\cos\theta)$ in Eq.\ (\ref{f}) are highly-oscillatory,
which makes it difficult the calculate scattering amplitude $f(\theta)$. Our solution is to cut-off the summation
at $\ell_b$, and let
\begin{equation}
    f(\theta)=\sum_{\ell=0}^{\ell_b}f^{\mathrm{WKB}}_\ell(\theta)-\sum_{\ell=0}^{\ell_b}f^{\mathrm{Born}}_\ell(\theta)+f^{\mathrm{Born}}(\theta), \label{f_app}
\end{equation}
which means we use the Born approxiamtion results of $f_\ell$'s for $\ell>\ell_b$ components \cite{Lin_2023}. Here,
\begin{equation}
    f^{\mathrm{Born}}(\theta)=-\frac{2\mu}{q}\int_0^\infty rU(r)\sin qr\d r,
\end{equation}
and \cite{wu2014quantum}
\begin{equation}
    f^{\mathrm{Born}}_\ell(\theta)=-\frac{\pi(2\ell+1)}{2k}P_\ell(\cos\theta)\int_0^\infty rU(r)J_{\ell+\frac{1}{2}}^2(kr)\d r.
\label{f_l_born}
\end{equation}
The integration containing a Bessel function in Eq.\ (\ref{f_l_born}) has an analytical form:
\begin{equation}
    \begin{aligned}
        &\int_0^\infty rU(r)J_{\ell+1/2}^2(kr)\d r=\frac{(2k)^{2\ell+1}\lambda_0^{2\ell+2}}{\pi}\\
        &\times B(\ell+1,\ell+1)_2F_1(\ell+1,\ell+1;2\ell+2;-4k^2\lambda_0^2),
    \end{aligned}
\end{equation}
where $_aF_b$ is the generalized hypergeometic function of order $a$, $b$.
Eq.\ (\ref{f_app}) becomes exact when $\ell_b$ is large enough such that $\delta_{\ell_b}^{\mathrm{WKB}}\simeq\delta_{\ell_b}^{\mathrm{Born}}$.

\section{Differential Cross-section due to ion-ion  collision \label{sec:Applicability-of-the}}
In this section, we generalize the original Bohr's criterion, which is derived based on
small angle Coulomb scattering, to a more general criterion that applies to arbitrary potentials and deflection angles.
This criterion is then benchmarked for DCS under repulsive Yukawa potential by numerical methods.

\subsection{The Generalized Bohr Criterion}

In 1948, N. Bohr proposed a famous criterion for the validity of the
classical orbital picture, now known as the Bohr criterion \cite{bohr1948penetration},
which is \cite{lindhard1965influence}
\begin{equation}
    \frac{1}{\mu v} \left| \frac{\d}{\d b}\theta^{-1} \right| \ll1,\label{bohr0}
\end{equation}
where $b$ is the collision parameter. For a small
angle Coulomb scattering ($\theta\ll1$), it becomes
\begin{equation}
    \frac{2Z_{1}Z_{2}}{ v}=\frac{2b_{0}}{\lambdabar}\equiv\kappa\gg1,\label{bohr}
\end{equation}
which means that the collision radius $b_0$ should be much larger than the
de Broglie wave length $\lambdabar=1/\mu v$. The Bohr's
formula Eq.\ (\ref{bohr}) implies that the quantum effect appears at high velocity ($v>2|Z_1Z_2|$) case for Coulomb scattering,
as is pointed out by Bounderup \cite{bonderup1981penetration}, this
is because the equation requires that $b_{0}\gg\lambdabar$,
and $b_{0}$ is proportional to $v^{-2}$ whereas $\lambdabar$ is
proportional to $v^{-1}$. 

The original Bohr's criteron (\ref{bohr}) is independent of the collisional parameter $b$.
Following Bohr's idea, generally, the following expression from Eq.\ (\ref{bohr0})
\begin{equation}
    \kappa\gg 2 \left| b_{0}\frac{\d}{\d b}\frac{1}{\theta} \right| \label{bohrgeneral}
\end{equation}
can be regarded as the criterion for arbitrary binary collisions with any
form of interactions, which should be dependent on the collisonal parameter. 
For example, for small angle scatterings with repulsive Yukawa potentials ($\sim\mathrm{e}^{-r/\lambda_{0}}/r$), 
the deflection angle can be approximated by \cite{sigmund2006particle}
\begin{equation}
    \theta(b)=\frac{2b_{0}}{\lambda_{0}}K_{1}\left(\frac{b}{\lambda_{0}}\right),
\end{equation}
with which Eq.(\ref{bohrgeneral}) (called as  the general Bohr criterion (GBC) in the following) reduces
to \cite{sigmund2006particle}
\begin{equation}
    \kappa\gg\lambda_{0}\frac{\d}{\d b}K_{1}^{-1}\left(\frac{b}{\lambda_{0}}\right).\label{lindhard}
\end{equation}
where $K_{\nu}(x)$ is the $\nu$'s order modified Bessel function
of the second kind. For collisions with not very low velocity, another approximation
was proposed in Ref.\ \onlinecite{mora2020coulomb}:
\begin{equation}
    \theta(b)=2\arctan\left[\frac{b_{0}}{\lambda_{0}}K_{1}\left(\frac{b}{\lambda_{0}}\right)\right],\label{yukawa_approx}
\end{equation}
which works well when $v \gtrsim 10^{-4}$ for ion-ion collision as found by us. When $b \ll \lambda_{0}$, $\frac{b_{0}}{\lambda_{0}}K_{1}\left(\frac{b}{\lambda_{0}}\right)$ becomes 
$\frac{b_{0}}{b}$. In this case $\theta(b)=2\arctan\left[\frac{b_{0}}{b}\right]$, which is just the result of strict Coulomb scattering.

\subsection{Numerical Benchmark for DCS}

\subsubsection{Results of the Generalized Bohr's Criterion}

\begin{figure}
    
    \includegraphics[scale=0.55]{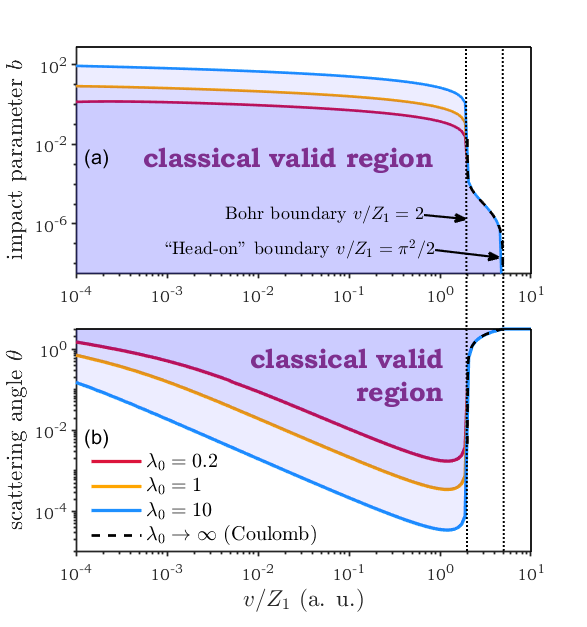}
    \caption{(a) The critical value of collision parameter ($b_{\mathrm{c}}$) at different $v$, beyond
        which the classical orbital picture is no longer valid according to
        the Bohr criterion Eq.\ (\ref{bohrgeneral}).
        (b) The deflection angle $\theta$ corresponding to the critical collision parameter $b_{c}$ at different $v$. 
         In both panels, $Z_1=2$ is fixed, and $\lambda_{0}$ is equal to 0.2, 1.0, and 10.0, respectively.}

    \label{bc}
\end{figure}

Here, by means of the numerical method mentioned in the last section,
we calculated the right-hand-side of Eq.\ (\ref{bohrgeneral}) as a
function of collision parameter $b$ for the scattering process 
in DT plasmas. The target particle is assumed to be an effective
particle with $Z_{2}=1$ and $m_{2}=2.5m_{p}$, where $m_{p}$ is
the mass of a proton. For simplicity, $m_{1}=2Z_{1}m_{p}$ is chosen. 
In Fig.\ \ref{bc} (a)  the y-axis is $b/\lambda_0$, and the three lines are the 
classical valid boundary $b=b_c(v)$ obtained by solving the equation
\begin{equation}
    \kappa=2 b_{0}\left| \frac{\d}{\d b}\frac{1}{\theta(b_c)}\right|,\label{gbc_eq}
\end{equation}
at three different screening lengths and beyond which the classical orbital picture is no longer valid.  
Fig.\ \ref{bc} (b) plotted the deflection angle boundary $\theta=\theta_c(v)$ calculated with the corresponding collision parameter in Fig.\ \ref{bc} (a). 
It should be mentioned that Eq.\ (\ref{gbc_eq}) is not the real form of GBC since later we will find that GBC is too strong to determine the range of $b$ or $\theta$ 
where the classical meachnics can be applied under the Yukawa potential.
Four behaviours can be seen from Fig.\ \ref{bc}:
\begin{enumerate}
    \item When $v<2|Z_1Z_2|$, $\theta_c$  ($b_c$) decreases(increases) as $\lambda_0$ increases, and becomes
           0 ($\infty$) for Coulomb potential when $\lambda_0\rightarrow\infty$.

    \item There are two velocity boundaries depending on the impact parameter, 
           below which the classical picture is valid. The fisrt ($v=2|Z_1Z_2|$) is refered to as
           the ``Bohr boundary'', while the second ($v=\pi^2|Z_1Z_2|/2$) the ``head-on boundary'', 
           since $\theta_c$ at this boundary is equal to $\pi$. 
           When $v > \pi^2|Z_1Z_2|/2$ no solution of $b$ for Eq.\ (\ref{gbc_eq}) can be found, which means that the classical picture is invalid for this region.
    
    \item When $v$ changes from $10^{-4}$ to $2|Z_1 Z_2|$, $b_{c}(v) / \lambda_0$ is almost independent of $\lambda_0$,
    while $\theta_c$ decreases as $\lambda_0$ increases.  
    \item When $v > 2|Z_1 Z_2|$, both $\theta_c$ and $b_{c}$ are independent upon $\lambda_0$, 
          and the $b_c$'s for both Coulomb potential and Yukawa potentials 
          are almost coincide.
  
\end{enumerate}

We will explain these behaviors one by one as follows:
\begin{enumerate}
\item When the impact parameter is large, the strength of Yukawa potential decreases rapidly,
      while the strength of Coulomb potential decreases slowly. Specifically,
      when $b\gg\lambda_0$, Eq.\ (\ref{yukawa_approx}) reduce to 
      \begin{equation}
        \theta(b)=2\arctan\left[\frac{b_{0}}{\lambda_{0}}\sqrt{\frac{\pi\lambda_0}{2b}}\exp\left(-\frac{b}{\lambda_{0}}\right)\right],
      \end{equation}
      which is very different from the Coulomb scattering. 
      This means that both $b_{\mathrm{c}}$ and $theta_{\mathrm{c}}$ are dependent upon $\lambda_{0}$ in the case.
\item The ``Bohr boundary'' $v=2|Z_1 Z_2|$ is exactly what Bohr's original criterion predicted,
    which is obtained under small angle approximation. As the scattering angle increases, the 
    boundary becomes larger, and reach to the ``head-on'' boundary when $\theta_c=\pi$.
We calculate the head-on boundary here. Solving the boundary equation 
\begin{equation}
\frac{1}{\mu v} \left| \frac{\d}{\d b}\theta^{-1} \right| =1 \label{boundary} 
\end{equation}
in oreder to obtain $b_c$ for Yukawa potential, we get
\begin{equation}
    \frac{b_{0}\left[K_{0}\left({\displaystyle \frac{b_{c}}{\lambda_{0}}}\right)+K_{2}\left({\displaystyle 
    \frac{b_{c}}{\lambda_{0}}}\right)\right]}{4k\lambda_{0}^{2}\left(1+{\displaystyle \frac{b_{1}^{2}}{\lambda_{0}^{2}}}\right)
    \arctan{\displaystyle \left(\frac{b_{1}}{\lambda_{0}}\right)}^{2}}=1, \label{yukawa_cri}
\end{equation}
where
\begin{equation}
    b_1\equiv \frac{Z_1Z_2}{\mu v^2}K_1\left(\frac{b_c}{\lambda_0}\right).
\end{equation}
For head-on collision $b$ should be very small so that $b_c\ll\lambda_0$ and $b_1\simeq\lambda_0 b_0/b_c$,
and by applying $K_0(x)\simeq -\ln x$, $K_1(x)\simeq x^{-1}$, and $K_2(x)\simeq 2x^{-2}$, we obtain
\begin{equation}
    \frac{b_{0}}{2k\left(b_{0}^{2}+b_{c}^{2}\right)\arctan\left(b_{0}/b_{c}\right)^{2}}=1. \label{coul_cri}
\end{equation}
Numerical calculation also shows that $b_0^2\gg b_c^2$ is generally true when $v>2|Z_1Z_2|$, hence Eq.\ (\ref{coul_cri}) reduces to
\begin{equation}
    \frac{v}{|Z_1Z_2|}=2\arctan\left(\frac{b_0}{b_c}\right)^{2}. 
\end{equation}
Since $\arctan\left(b_{0}/b_{c}\right)^{2}\leq \pi^2/4$, $v$ should not be more than $|Z_1Z_2|\pi^2/2$,
which explains the head-on boundary.

\item
Fig.\ \ref{bc} (a) tells us that when $v<2|Z_1Z_2|$, $b_c\gg\lambda_0$ or $b_c\simeq\lambda_0$,
which means $b_1$ may be greater or smaller than $\lambda_0$. If $b_1<\lambda_0$, we have $\arctan\left(b_{1}/b_{c}\right)\simeq b_{1}/b_{c}$
and $1+{\displaystyle b_{1}^{2}/\lambda_{0}^{2}}\simeq 1$, then Eq.\ (\ref{yukawa_cri}) reduces to
\begin{equation}
    \frac{b_{0}}{4kb_{1}^{2}}\left[K_{0}\left({\displaystyle \frac{b_{c}}{\lambda_{0}}}\right)+K_{2}\left(\frac{b_{c}}{\lambda_{0}}\right)\right]=1.
\end{equation}
Similarily, if $b_1>\lambda_0$, then $\arctan(b_1/\lambda_0)$ is ranging from $\pi/4$ to $\pi/2$, and becomes weakly dependent on $b_1/\lambda_0$. 
Thus we approximately let $\arctan(b_1/\lambda_0)$ equal to a constant $C$. Eq.\ (\ref{yukawa_cri}) thus becomes
\begin{equation}
    \frac{b_{0}}{4kb_{1}^{2}C^{2}}\left[K_{0}\left({\displaystyle \frac{b_{c}}{\lambda_{0}}}\right)+K_{2}\left({\displaystyle \frac{b_{c}}{\lambda_{0}}}\right)\right]=1.
\end{equation}
Obviously in these two cases the solution is dependent on $b_c/\lambda_0$, not $\lambda_0$ if we notice that 
$\frac{b_{0}}{4kb_{1}^{2}}=\frac{v}{4|Z_1 Z_2| K_{0}^2\left({\frac{b_{c}}{\lambda_{0}}} \right)}$ in the above two equations. 
This explains the observed second behavior. 
Also thus indicates that $b_c$ is independent on reduced mass $\mu$.

\item
The boundary value $b_c$ and $\theta_c$ for $v>2|Z_1Z_2|$ is determined by Eq.\ (\ref{coul_cri}), which is independent of
$\lambda_0$. This is also the boundary of $v$ for large angle Coulomb scattering 
since $\theta (b)$  for Debye potential becomes that for Coulomb potential in the case.
\end{enumerate}
Now it is necessary for us to further discuss the behavior of $\theta_c$ in Fig.\ \ref{bc} (b.) 
From the figure we found that, 
$\theta_c$ depends on $\lambda_0$ and decreases with $v$ when  $v < 2|Z_1 Z_2|$, 
and $\theta_c$ rapidly rises up to $\pi$ for higher $v$. 
This is easy to see if Eq.\ (\ref{yukawa_approx}) is noticed. 
The equation means that the scattering angle is a function of $\lambda_0$ 
when $b \ll \lambda_0$ is not satisfied (since $K_1(x)\simeq x^{-1}$ for small $x$), 
which correspounds to the case for $v < 2|Z_1 Z_2|$ from Fig.\ \ref{bc} (a). 
For $v > 2|Z_1 Z_2|$ due to that $b_{c}$ is very small or even smaller than $b_0$, 
the corresponding scattering angle becomes large and even close to $\pi$. 
Besides this, $\theta_c$ is decreasing with $\lambda_0$ rising when  $v < 2|Z_1 Z_2|$, 
which is understandble from the equation since 
$b_{c} / \lambda_0$ is almost independent on $\lambda_0$ and decreases with $v$ in the case. 
In addition, $b_0$ rapidly decreases with $v$ increasing, 
which results in the reducing of $\theta_c$ 
at the same time according to the equation when $v < 2|Z_1 Z_2|$.

\begin{figure}
    
    \includegraphics[scale=0.55]{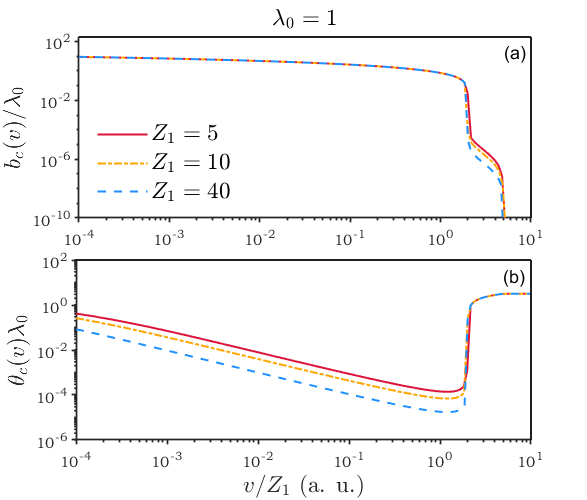}
    \caption{(a) The critical value of collision parameter ($b_{\mathrm{c}}$) at different $v/Z_1$.  
    (b) The deflection angle $\theta$ corresponding to the critical collision parameter $b_{c}$ at different $v/Z_1$.
    in both panels $\lambda_0$ is fixed to be 1.0, and three different projectile charge states  with $Z_1$ equal to 5, 10 and 40 are chosen.
    }
    \label{bc_Z}
\end{figure}

The above analysises are further embodied in Fig.\ \ref{bc_Z} (a) and Fig.\ \ref{bc_Z} (b), where  the evolutions of
$b_{c} / \lambda_0$ and $\theta_c$ with $v / Z_1$ are shown, respectively 
when $Z_1=$5, 10, and 40 for $\lambda_0$ fixed to be 1.0. Here the explanation of the figure will  be given no longer.  Similar results with the figure
for many other $Z_1$ and $\lambda_0$ have also been obtained and not shown here.

\subsubsection{Comparison of Classical and Quantum DCS}

\begin{figure*}
  
    \includegraphics[width=0.24\textwidth]{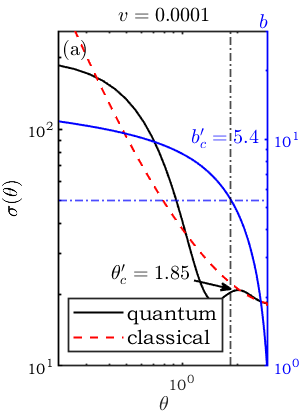}
    \includegraphics[width=0.24\textwidth]{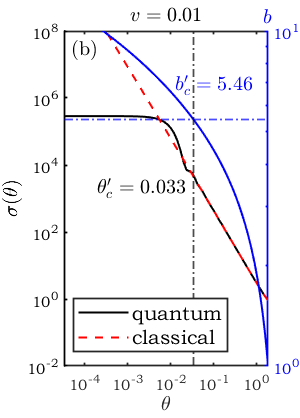}
    \includegraphics[width=0.24\textwidth]{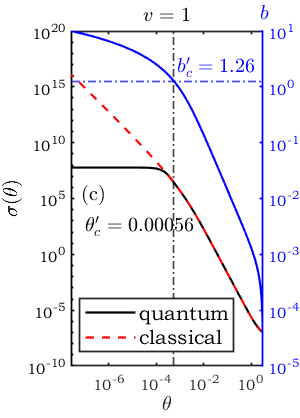}
    \includegraphics[width=0.24\textwidth]{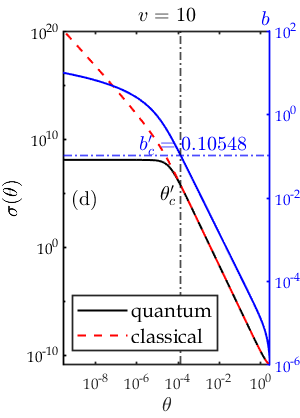}
  
    \caption{Variations of impact paramter $b$, classical and quantum differential cross sections  with scattering angle $\theta$ 
    when $v=10^{-4}$, $10^{-2}$, $1.0$, and $10.0$. Here $Z_1=2.0$ and $\lambda_0=1.0$ are chosen. 
    Both $b_{c}'$ and corresponding angle $\theta_{c}'$ are marked, where
        $b_{c}'$ is where the classical and quantum differential cross-section
        starting to disagree (about 2\% deviation). }
    \label{bcthetac}
\end{figure*}

So far we have  presented the range of $b$ or $\theta$ 
where the classical meachnics can be applied under the Yukawa potential according to Eq.\ (\ref{gbc_eq}) related to Bohr criterion. 
The reliability of the result need be tested by the comparison of classical and quantum DCS. For this aim 
we now first take a look at the DCS by both classical and qunatum methods. 
In order to calculate the DCS by means of Eq.\ (\ref{qdc}), we take the result of PW method 
as the exact quantum mechanical phase shift if there is a substantial difference 
between the PW and the WKB results.  Fig.\ \ref{bcthetac} shows both the results of classical 
and quantum DCS  as a function of scattering angle $\theta$ when $v=10^{-4}$, 
$10^{-2}$, $1.0$, and $10.0$. Here $Z_1=2.0$ and $\lambda_0=1.0$ are chosen. The 
relevant variations of impact paramter $b$ with $\theta$ are also plotted. 
Obviously the DCS calculated via the two methods are almost identical for 
relatively large deflection angle $\theta$, and their difference appears 
obviously with the decreasing of $\theta$ or corresponding  to bigger impact 
parameter. Morever, the quantum 
result converges to a finite value as $\theta\rightarrow0$ while the classical 
result does not, which means that the classical total cross section is divergent 
while the quantum result is not.
\begin{figure}
    
    \begin{centering}
        \includegraphics[width=\columnwidth]{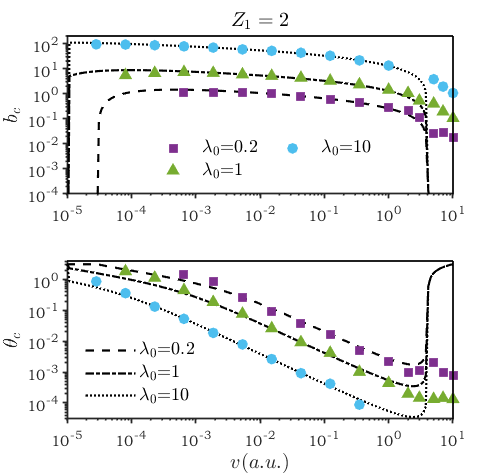}
        \par\end{centering}
    \caption{Variations of $b_{c}'$ / $\theta_{c}'$ (markered lines)  and $b_{c}$ / $\theta_{c}$ (unmarkered lines) as a function of $v$ 
    for $\lambda_0=0.2$, 1, and 10 when $Z_1=2.0$}

    \label{bcprime}
\end{figure}
We then select a specific deflection angle $\theta_{c}'$, at
which the quantum and classical results just begin to deviate about
2\%, and set the corresponding collision parameter as the
actual critical collision parameter $b_{c}'$ which may be different from $b_c$.  For the case of $Z_1=2$, $\lambda_0=0.2$, 1.0, and 10.0 
all the $b_{c}'$ and $b_c$ are plotted in Fig.\ \ref{bcprime}. Three features can be seen from the figures:
\begin{enumerate}
    \item When  $v$ is smaller than $10^{-3}$, $b_{c}'$ and $b_c$ does not match very well. 
    \item When $v$ is between $10^{-3}$ to $1$,  $b_{c}'$ and $b_c$ are close to each other.
    \item When $v$ is approaching or beyond $2|Z_1Z_2|$, $b_{c}'$ and $b_c$ become to  match no longer.
\end{enumerate}
And so does $\theta_c$ and $\theta_c'$. Again, we explain them one by one here:

\begin{enumerate}
    \item When  $v$ is very small ($ < 10^{-3}$ ), the quantum
    and classical DCS is different in the most range of $\theta$, it
    is difficult to obtain a critical deflection angle $\theta_{c}'$
    and thus the critical collision parameter $b_{c}'$.
    In this case we think the quantum wave character plays  its dominant role to determine the DCS
    and the concept of classical impact parameter, which was used in Bohr criterion, is invalid. 
    The reason is that the corresponding de Broglie wave length $\lambdabar > 3.54$
    for the collision system, which is smaller than the screening lengths $\lambda_0$ in the figure. 
    This makes $k \lambda_0 =\frac {\lambda_0} {\lambdabar}  <1 $. In other words, 
    only very few number of partial wave $l$ has contribution to the quantum DCS.   
 
    \item The second feature confirms that the physical interpretation
    of $\theta_c$ is indeed the critical value of scattering angle below which the quantum and 
    classical DCS starting to deviate. Meanwhile the GBC will results in a  range of much smaller $b$ or much bigger $\theta$ 
    that the classical mechanics can be applied.  In other words, in the case the GBC is too strong to get a proper range wher the clasical picture is reliable. 
    \item We think that in the case both the GBC and its weaker form Eq.\ (\ref{gbc_eq}) are invalid to give a proper value of $\theta$ 
    below which the classical mechanics does not work though we are not clear to this. In fact only in the range of very small $\theta$ or enough big $b$ 
    the classical and quantum DCSs become different. This has something to do with the failure of  Bohr criterion for the scattering under Coulomb potential.
     It is well known 
     that the two DCSs by both classical and quantum mechanics are always the same for Coulomb potential \cite{landau2013quantum} 
     while the original Bohr criterion suggest that the classical picture fails when $ v \ll 2|Z_1 Z_2|$. 
     The difference between Coulomb and Yukawa potentials occurs when $r > \lambda_0$. In scattering it corresponds to enough large $b$ or small $\theta$,
      which is just the range where the classical picture does not work.      
\end{enumerate}
Furthermore, for the case of high enough velocity  the discrepancy between the two DCSs by both classical and quantum mechanics can be seen from 
the related analytic expressions. The corresponding classical Yukawa DCS is\cite{mora2020coulomb}
\begin{equation}
    \sigma(\theta)=\sigma^{C}(\theta)\frac{xK_{1}^3(x)}{\left|\d K_{1}/\d x\right|},
\end{equation}
which reduces to the Coulomb cross-section $\sigma^{C}(\theta)$ when $x\equiv b/\lambda_0$ is small enough. 
However, for ion-ion collision in high speed $b_0 \ll \lambda_0$ 
so that the corresponding $\theta$ for such $x$ covers a large range from small angle to $\pi$
For quantum mechanical Born approxiamtion Eq.\ (\ref{born_app}) also usually reduces to  $\sigma^{C}(\theta)$ only if $\theta$ is not very small. 
All these mean that the discrepancy between the two DCSs by both classical and quantum mechanics only appears for very small $theta$, as shown in the above two figures.
This further varifies that the GBC fails in the case.  

\section{Results of the Transport Cross-Section}

The transport cross-section is what really matters in a collision-based
method \cite{sigmund1982kinetic}. In the section we first analyse the asymptotic behaviour of the integrands for 
classical and quantum TCS in low and high velocity limits, and then test
the analysis by numerical results of both classical and quantum results of TCS.

\subsection{Low Velocity Case}

\begin{figure*}

    \includegraphics[width=0.24\textwidth]{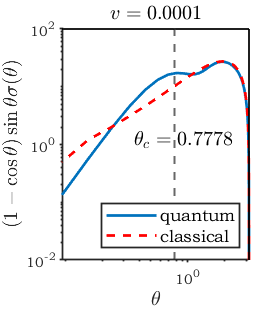}
    \includegraphics[width=0.24\textwidth]{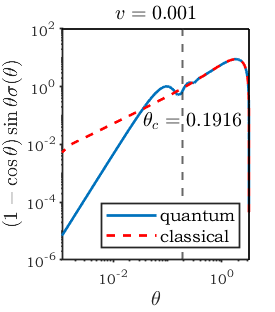}
    \includegraphics[width=0.24\textwidth]{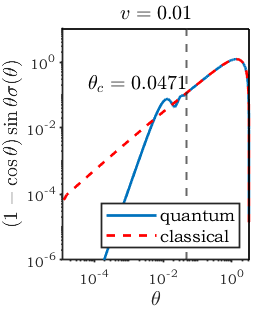}
    \includegraphics[width=0.24\textwidth]{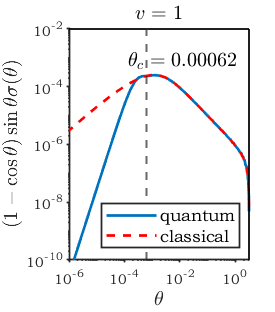}

    \caption{The integrands of TCS ($\lambda_0=1$) with $Z_1=2$ and  
     $v=10^{-4},10^{-3}, 10^{-2}$ and 1, where the solid-lines are quantum mechanical results
    and dashed-lines are classical with the corresponding $\theta_{c}$ marked.}  
    \label{sigma_tr_itgd_low}
\end{figure*}

In the classical calculations,
the transport cross-section depends not only on the differential cross section
$\sigma(\theta,v)$, but also the factor $\left(1-\cos\theta\right)\sin\theta$,
according to Eq.\ (\ref{ctrdc}). We calculate integrand of Eq.\ (\ref{ctrdc}) with 
the same parameter used in the previous section, the result plotted in
Fig.\ \ref{sigma_tr_itgd_low}, where the critical deflection angle $\theta_{c}$
is the corresponding angle to the critical collision parameter presented
in the last section. As we can see, at very low collision velocity ($v=0.0001$),
the integrand of classical and quantum TCS is quite different with $\theta_{c}$ much bigger than $\theta$, and
the Bohr's criterion does not work well at such low velocity as explained in the before.

\subsection{Middle Velocity Case}

For medium velocities ($v=0.001,0.01$, and even $v=1.0$), the discrepancy of classical
and quantum integrands of DCS become large only  in small-$\theta$ region though
they are suppressed by the $(1-\cos\theta)\sin\theta$ factor. At higher $\theta$ 
the integrands thus approximately coincide with each other. Notice
that the integrand in the range $0\sim\theta_{c}$ only contribute
1.79\%, 0.13\%, and 0.08\% to the final results of TCS for $v=0.001$, $v=0.01$ and $v=1$ respectively, and this
is the reason why the quantum and classical TCS results are almost coincide with each other
even though the classical DCS is divergent while quantum DCS is not. Only in this case the Bounderup's view is valid. 

By the way here for the above two cases we only  show the results for $Z_1=2.0$ and $\lambda_0=1$. 
For other $Z_1$ and $\lambda_0$ similar results with the figure have also been found and not shown here.

\subsection{High Velocity case}
At high collision velocity, the Born approximation of the
transport cross-section gives
\begin{equation}
    \begin{aligned}\sigma_{\mathrm{tr}}(v)= & 2\pi\int_{0}^{\pi}\left(1-\cos\theta\right)\sin\theta\sigma_{\mathrm{Born}}^{\mathrm{Y}}\left(\theta\right)\d\theta                           \\
               =                        & 2\pi b_{0}^{2}\left[\ln\left(1+4\mu^{2}v^{2}\lambda_{0}^{2}\right)-\frac{4\mu^{2}v^{2}\lambda_{0}^{2}}{1+4\mu^{2}v^{2}\lambda_{0}^{2}}\right] \\
               \simeq                   & 4\pi b_{0}^{2}\ln\left(\mu v\lambda_{0}\right),
    \end{aligned}
    \label{highvsigmatrq}
\end{equation}
where in the last line we have assumed $\mu v\lambda\gg1$. On the
other hand, the classical calculation \cite{mora2020coulomb} predicts that 
when $\mu v\lambda/\left|Z_{1}Z_{2}\right|\gg1$,
\begin{equation}
    \sigma_{\mathrm{tr}}(v)\simeq4\pi b_{0}^{2}\ln\left(\frac{2\lambda_{0}}{b_{0}}\mathrm{e}^{-\gamma-\frac{1}{2}}\right).\label{highbsigmatrc}
\end{equation}
One find that Eq.\ (\ref{highvsigmatrq}) and Eq.\ (\ref{highbsigmatrc}) only coincident when
$v/\left|Z_{1}Z_{2}\right|\simeq \mathrm{e}^{\gamma+\frac{1}{2}}/2$, which means that when
\begin{equation}
v>1.468\left|Z_{1}Z_{2}\right|,
\end{equation}
the classical and quantum TCS should be different.

\begin{figure}

    \begin{centering}

    \includegraphics[width=0.45\textwidth]{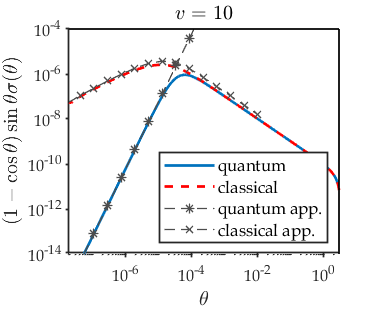}
    \includegraphics[width=0.45\textwidth]{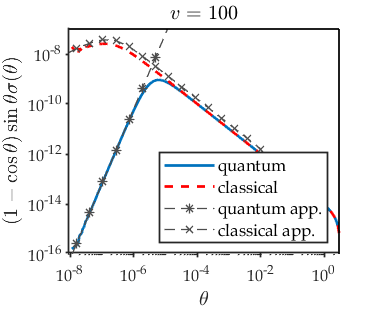}\par
    \end{centering}
    \caption{The integrands of TCS with $\lambda_0=1$ and $Z_1=2$ at high
    collision velocity ($v=10$, $100$), where the solid-lines are quantum mechanical results
    and dashed-lines are classical. The quantum and classical approximation curves
    are calculated via Eq.\ (\ref{quantumtheta0}) and Eq.\ (\ref{classtheta0}) respectively. }
    \label{sigma_tr_itgd_high}
\end{figure}

To be specific, for high velocity collision, we can concentrate on
the limit where $\theta\rightarrow0$. At this limit, according to Eq.\ (\ref{born_app})
the Born approximation DCS
converge to a constant:
\begin{equation}
    \sigma(\theta\rightarrow0)=4Z_{1}^{2}Z_{2}^{2}\mu^{2}\lambda_{0}^{2}. \label{quantumtheta0}
\end{equation}
Recall that the classical Yukawa DCS is \cite{mora2020coulomb}
\begin{equation}
    \sigma(\theta)=\sigma^{C}(\theta)\frac{xK_{1}^3(x)}{\left|\d K_{1}/\d x\right|},
\end{equation}
where $x=b/\lambda_{0}$. Noticing that $K_n(x\gg 1)\simeq\mathrm{e}^{-x}\sqrt{\pi/2x}$, we find that for $\theta\rightarrow0$,
\begin{equation}
    \begin{aligned}
    \sigma(\theta\rightarrow0) &=\frac{\pi}{2}\sigma^{C}(\theta)\mathrm{e}^{-2x}\\
    &\simeq\frac{\pi}{2}\sigma^{C}(\theta)\exp\left[-LW\left(\frac{4\pi b_{0}^{2}}{\theta^{2}\lambda_{0}^{2}}\right)\right],
    \label{classtheta0}
    \end{aligned}
\end{equation}
where $LW(x)$ is the Lamberg W-function \cite{wright1959solution}. Therefore, the asymptotic behaviors of classical and quantum high-velocity 
DCS are quite different for very small angle scattering. However, in the following we will see that in the case the region of  very small angle 
have an important contribution to the TCS although for other range of $\theta$ the quantum DCS is close to the classical one as mentioned in the 
endo of last section.  
   In Fig.\ \ref{sigma_tr_itgd_high} the integrands in the total ranges of angles are plotted 
  for two different high $v$ with $v=10$ and $100$ at $Z_1=2$ and $\lambda_0=1$. 
It is easy to see that, despite that the Bohr criterion is 
absolutely violated at such high speed, there is still a large portion of the $\theta$ range 
where the classical and quantum DCS coincide.
 However, the contribution from the part of small angle to the TCS is important in this case. 
 By integration it is found that contributions of the part with $\theta < 10^{-3}$  are about 26\% of the total quantum TCS,  
and the discrepancy of DCS at $v=10$ and $100$ are 8.2\% and 32\% respectively. 

\subsection{Numerical Results of the TCS}
\begin{figure*}
    \includegraphics[width=\textwidth]{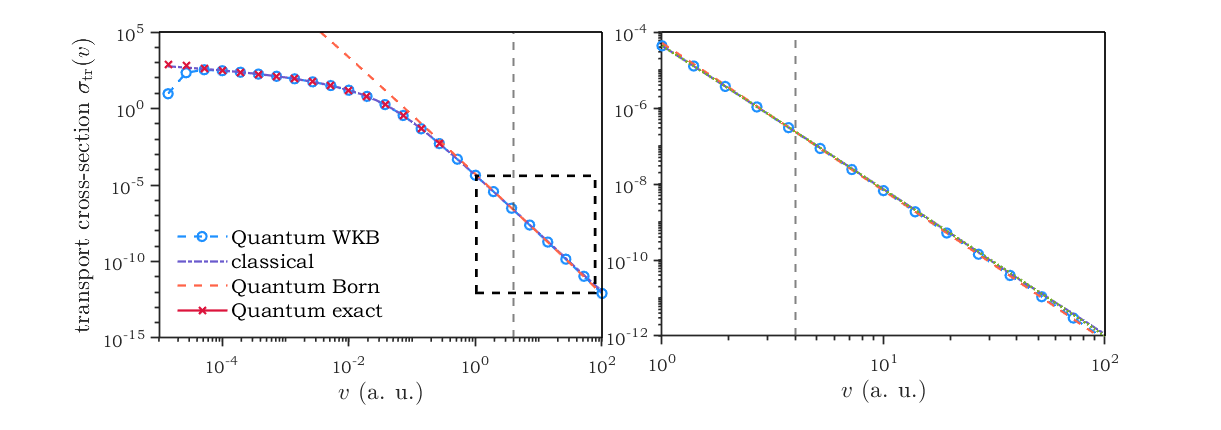}

    \caption{Transport cross-sections of ion-ion collision under Yukawa potential
    ($\lambda_0=1$ and $Z_1=2$) calculated via WKB approx., classical, Born approx., and exact partial
        wave method respectively. The dashed-line marks $v=2|Z_1Z_2|=4$. The discrepancy in high-velocity range can be seen more clearly in the right panel.}
    \label{sigmatr_qvc}
\end{figure*}

\begin{figure}
    \begin{centering}
        \includegraphics[width=0.45\columnwidth]{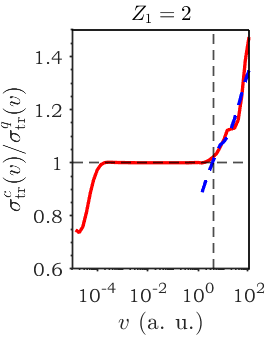}
        \includegraphics[width=0.45\columnwidth]{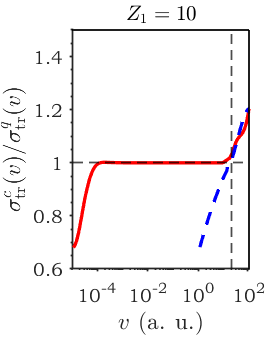}
        \par\end{centering}
    \caption{(Color online) The ratio of classical TCS to quantum one with $Z_1=2$ (a) and $10$ (b) at $\lambda_0=1$.
    The red solid-line is calculated by numerical methods described in the context, and the blue dashed-line
    is the result from the approximation formulas Eq.\ (\ref{highbsigmatrc}) and Eq.\ (\ref{highvsigmatrq}). }
    \label{rate}
\end{figure}
In this subsection, the exact value of quantum and classical TCS are
calculated by means of the aforementioned numerical methods. Notice
that the classical TCS can be reduced to
\begin{equation}
    \sigma_{\mathrm{tr}}(v)=2\pi\int_{0}^{\infty}\left[1-\cos\theta(v,b)\right]b\d b,
\end{equation}
according to Eq.\ (\ref{cdc}). The result is presented in Fig.\ \ref{sigmatr_qvc},
from which one can see that the quantum TCS and the classical TCS
coincide in the range of $v$ between $10^{-3}$ and about $Z_{1}Z_{2}$. In smaller $v$
region, the WKB approximation result is obviously incorrect, let alone
the Born approximation result. Hence, we choose the result of exact
partial wave method to be the real quantum TCS in the case. In the larger $v$
region, the classical and quantum (WKB) results deviate for each other
a little bit, and the WKB result almost coincide with the Born approximation.
This deviation takes places when $v>2|Z_{1}Z_{2}|$.
In Fig.\ \ref{rate}, we plotted the ratio of classical TCS to  
quantum TCS for two different projectiles ($Z_1=2,10$). For some other $Z_1$ and $\lambda_0$ similar results are also calculated and not shown here.
One can see from the figure that although the classical and quamtum TCS are very close to each other for a large range of $v$, 
their discrepancy is quite clear for very low and high $v$ regimes.
Part of the relevant reason  has been pointed out above. For very low velocity
collisions, only the lowest several levels of partial wave $\ell$'s are involved since the de Broglie wave length is long, 
the classical orbital picture is thus invalid for such a collison.  At high $v$ region 
the DCS for quantum and classical scattering of small angle is quite different, and the small angle scattering play a significant role to determine the TCS. 
This results in an obvious difference of TCS from classical and quantum mechanics.

\section{Conclusion}

In this paper, we examined the validity of classical mechanics to descibe the elastic ion-ion collision 
under repulsive Yukawa potential. Both results of classical and quantum DCSs are compared in detail as well as those of TCSs. 
The relevant results for the validity of classical mechanics are compared  by the generalized Bohr's criterion (GBC).
Our main conclusions are summarized as follows: 

\begin{enumerate}
    \item   
    For very low-velocity collisions quantum and classical DCS are quite different 
    in a large range of scattering angle. The reason is that the collision process is dominant by quantum wave effect in the case 
    so that only very few partial waves contribute to the DCS. 
    As velocity increases, the region that quantum and classical DCSs are different become smaller,which is towards to  more and more small angle. 
     This behavior agrees with the 
    prediction of the not so strict GBC when $v<2|Z_1Z_2|$.
    \item 
     For low-velocity collisions quantum and classical TCSs are obviously different since the corresponding DCSs are quite different 
     for a large range of scattering angle.
     As velocity increases, the discrepancy becomes negligible
      where the discrepancy between the corresponding DCSs in small angle has few affect to the TCS.
    However, when $v>2|Z_1Z_2|$, the discrepancy occurs again, since in such high velocity the small angle scattering becomes dominant, 
    which has an important affect upon the TCS.
 
    \item The GBC is too strong to get a proper range where the clasical picture is reliable. Its weaker form Eq.\ (\ref{gbc_eq})
    is invalid for very low velocity collision since the concept of impact parameter does not work in the case. 
    Meanwhile in middle velocity range the weaker form can predict a reliable range where the clasical picture is reliable. 
    The Bohr criterion does not work when $v>2|Z_1Z_2|$, which coincides with its  failure for Coulomb scattering for such velocity. 
   
\end{enumerate}

Anyhow, the difference between quantum and classical TCS has minor effects to the stopping power,
the classical picture of ion-ion collision under ICF parameter is justified.
Still, we hope that this paper serves as a reminder that the
idea that ion-ion collision is a purely classical process should not
be taken for granted but actually has deeper physical significance.

\section*{Acknowledgement}
This work was supported by the National Key R\&D Program of China under Grants (No. 2022YF1602500),
National Natural Science Foundation of China under Grants (No. 12075204, No 12274039), 
the Strategic Priority Research Program of Chinese Academy of Sciences (Grant No. XDA250050500),
the Shanghai Municipal Science and Technology Key Project (No. 22JC1401500), and IAEA Research Contract (No. 24243/R0).
D. Wu thanks the sponsorship from Yangyang Development Fund. B. He thanks Prof. P. Sigmund for the kind discussions. 


\bibliographystyle{unsrt}
\bibliography{ref.bib}

\end{document}